# RECTIFICATION MECHANISM IN DI-BLOCK OLIGOMER MOLECULAR DIODES

M.A. Kozhushner, and V.S. Posvyanskii
Institute of Chemical Physics RAS, Moscow, Russia

I.I. Oleynik
Department of Physics, University of South Florida, Tampa, FL

L. Yu
Department of Chemistry and James Franck Institute, The University of Chicago, Chicago, IL

**Abstract**

We investigated a mechanism of rectification in di-block oligomer diode molecules that have recently been synthesized and showed a pronounced asymmetry in the measured I-V spectrum. The observed rectification effect is due to the resonant nature of electron transfer in the system and localization properties of bound state wave functions of resonant states of the tunneling electron interacting with asymmetric molecule in an electric field. The asymmetry of the tunneling wave function is enhanced or weakened depending on the polarity of applied bias. The conceptually new theoretical approach, the Green's function theory of sub-barrier scattering, is able to provide a physically transparent explanation of this rectification effect based on the concept of the bound state spectrum of a tunneling electron. The theory predicts the characteristic features of the I-V spectrum in qualitative agreement with experiment.

In their pioneering paper [1] Aviram and Ratner proposed the idea of a molecular rectifier that contains donor and acceptor π-conjugated segments separated by an insulating σ-bonded segment of molecular wire. Several molecular rectifying systems have been synthesized in the past decade using Langmuir-Blodgett molecular assembly [2-6]. Attempts to provide experimental proof of molecular rectification were complicated by difficulty in establishing reproducible electrical contacts between metallic electrodes and a single molecule which resulted in uncontrollable interface rectification effects [7].

Recently, a new class of diode molecules has been synthesized based on di-block oligomer molecules [8,9]. These molecules consisting of thiophene and thiazole structural units, have shown a pronounced rectification effect as a result of built-in chemical asymmetry. Importantly, it was unambiguously shown that the rectification effect is an intrinsic property of di-block oligomer molecules, and not due to the molecule-electrode interfacial interactions. In addition, by synthesizing diode molecules with different terminal thiol groups, it has become possible to assemble the diode molecules between gold electrodes with pre-defined rectification direction [10].

In this letter we explain the mechanism of rectification in di-block oligomer diode molecules. We demonstrate that the observed asymmetry of current-voltage characteristics is the result of the resonant character of electron transport in molecular diodes and spatial asymmetry of the wave-function of a tunneling electron interacting with asymmetric molecule in an applied electric field. The asymmetry of the tunneling wave function is enhanced or weakened depending on the polarity of applied bias.

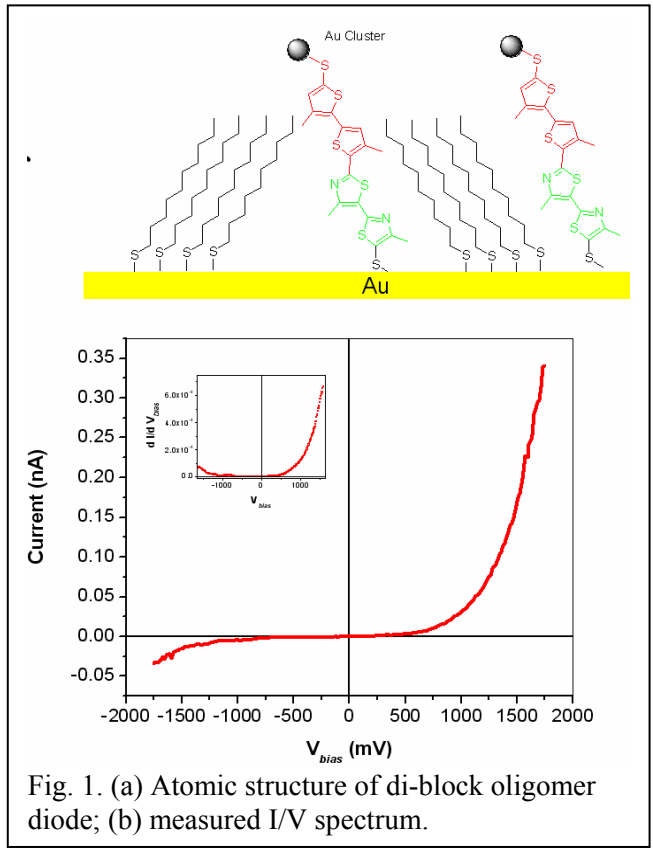

Fig. 1. (a) Atomic structure of di-block oligomer diode; (b) measured I/V spectrum.

The di-block oligomer molecule consists of two parts each containing an equal number of thiophene ($C_4N$) and thiazole ($C_3NS$) rings, see Fig. 1a. Two-terminal molecular circuits is produced by sequential assembly between two gold electrodes, and the current-voltage characteristics of an individual molecule is measured by scanning tunneling spectroscopy [8-10]. The characteristic feature of an I-V spectrum of molecular diodes is threshold voltage i.e. the absence of current below some value of applied bias. The threshold is observed for both polarities of the bias, see Fig. 1.b.

We have recently shown [11] that the presence of a threshold in the I/V spectrum is a distinct signature of resonant electron transport in a single molecule. In the electrode-molecule-electrode system the electron tunnels through the organic molecule at negative energies with respect to the vacuum level. An important question is concerned with the nature of the energy spectrum of the tunneling electron. Usually, the electron orbitals of a neutral molecule are calculated and used for interpretation of the transport mechanisms. In reality, the tunneling electron is an extra electron that interacts with the neutral molecule (with its electrons and nuclei) in the course of electron transition. Therefore, its energy spectrum at negative energies corresponds to the energy levels of the "negative ion", i.e. the bound energy spectrum of the system of one electron plus a neutral molecule [12].

In order to address this issue we have developed a theory of the energy spectrum of a tunneling electron in a molecular wire based on the concepts of sub-barrier scattering [13,14]. Depending on the properties of the energy spectrum and its position with respect to the Fermi levels of the electrodes, two fundamental mechanisms of electron transport through single molecules are possible. They are ordinary tunneling, when the energy levels of a tunneling electron are above the Fermi level of the negatively biased electrode, and resonant electron transfer, when part of the spectrum is within the energy interval $\varepsilon_F^l - |V| \leq \varepsilon_s \leq \varepsilon_F^l$. In particular, resonant electron transfer produces a threshold in applied voltage when the first lowest energy level is aligned with the Fermi energy of the left electrode $\varepsilon_F^l$, see Fig. 2. In Fig. 2 the positive bias is applied to the right electrode so that the electrons of the left electrode are transferred to the right electrode via the resonant energy levels of the negative ion.

In general, the energy spectrum of a neutral molecule is very different from the energy spectrum of an extra electron interacting with the molecule, i.e. the energy spectrum of the negative ion. Therefore, the traditional concepts of quantum chemistry such as HOMO and LUMO frequently used to interpret transport through molecules are not applicable to the analysis of electron transport through a single molecule. In addition, the electrons in an organic molecule are tightly bound and from the point of view of electronic structure it can be classified as an insulator. Consequently, the concepts of semiconductor physics including notions of n- and p-conductivity that assume the presence of free carriers are not applicable to the case of transport through single molecules.

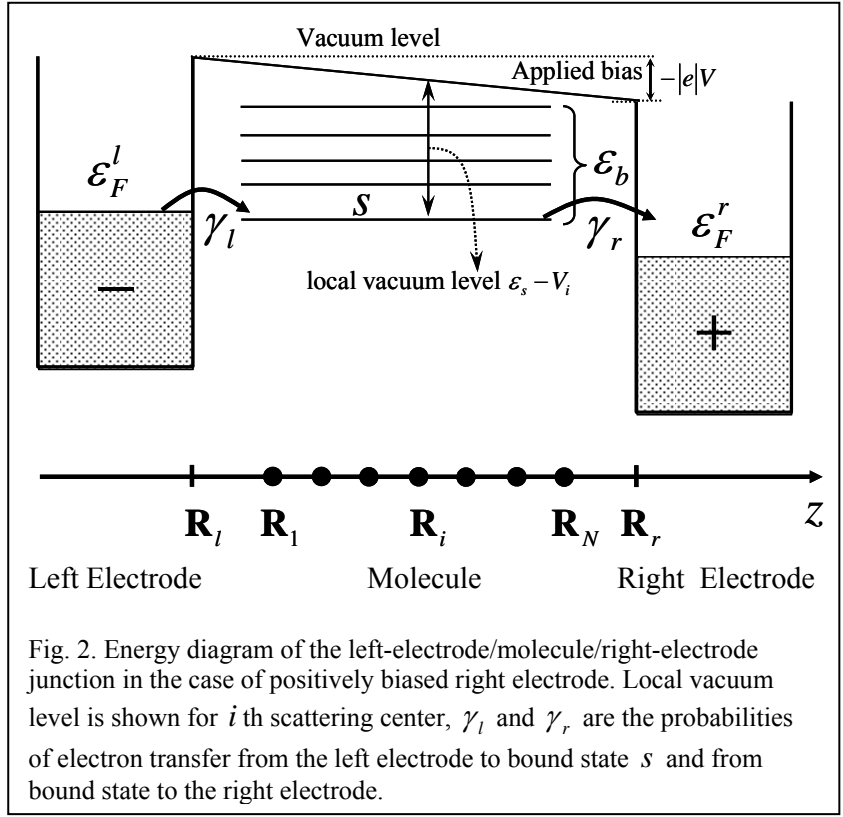

Fig. 2. Energy diagram of the left-electrode/molecule/right-electrode junction in the case of positively biased right electrode. Local vacuum level is shown for $i$ th scattering center, $\gamma_l$ and $\gamma_r$ are the probabilities of electron transfer from the left electrode to bound state $s$ and from bound state to the right electrode.

Taking into account the resonant character of electron transport in molecular diodes substantiated by the experimentally observed threshold in the I-V spectrum, one might ask the question what are the factors contributing to the asymmetry of the current with respect to the polarity of applied bias? From the standard viewpoint the difference in local electronic properties of two parts of the molecule can not result in an asymmetry of the current because both right and left ends of the molecule make contributions to the matrix element of the resonant transition that results in the symmetrical expression for the resonant current due to the resonant state $s$:

$$I_{res}(\varepsilon_s) = \frac{1}{2}\frac{\gamma_l \gamma_r}{\gamma_l + \gamma_r} \tag{1}$$

where $\gamma_{l,r} = 2\pi |\langle \psi_{l,r} |U| \psi_b \rangle|^2 \rho_{l,r}(\varepsilon_s)$ are the probabilities of electron transfer from the bound energy state $\psi_b$ to one of the states $\psi_l$ or $\psi_r$ of the continuous spectrum of the left or right electrode, $\gamma_s = \gamma_r + \gamma_l$ being the inverse lifetime of the resonant energy level; $U$ is the electronic potential at the electrode-molecule interface (we use the atomic system of units $\hbar = e = m_e = 1$). Expression (1) is symmetric with respect to the left and right electrodes and, consequently, will give approximately the same value of current at small applied voltages $\pm V$, $|V| \ll 1 \mathrm{V}$ even if there is some small dependence of the bound state wave function $\psi_b$ on the applied electric field $E$.

However, an important feature of molecular rectification is the resonant character of electron transfer through the diode molecule which results in a threshold voltage $V_{l,r}^{th} \sim 1$ eV. At such bias the spatial behavior of the bound state wave function of the tunneling electron becomes highly asymmetric because of a small penetration of the bound state wave function under a triangular potential barrier $U(z) = -Vz/d$ ($d$ is the distance between left and right electrodes). Then, the probability of electron transfer from the left electrode to the resonant bound state $\gamma_l$ is very small compared to that from the bound state to the right electrode $\gamma_r$, see Fig. 2. This creates a bottleneck at the left electrode for electron transfer at positive bias. For negative bias electrons are transferred from the right electrode to the left electrode and the bottleneck is created at the right electrode $\gamma_r \ll \gamma_l$.

In both cases, the resonant current acquires an asymmetry with respect to the polarity of an applied bias:

$$I(\varepsilon_s) = \begin{cases} 1/2\gamma_l, & V \geq V_l^{th} > 0 \\ 1/2\gamma_r, & V \leq V_r^{th} < 0 \end{cases}. \qquad (2)$$

Because the probabilities of electron transfer to the left and right electrodes are proportional to $|\psi_b(\mathbf{R}_{l,r})|^2$, the asymmetry in the current as the polarity of bias changes will be due to the inequality $\gamma_l \neq \gamma_r$. This is because chemically different parts of the molecule interact differently with the tunneling electron. However, the standard calculations of the wave-function of the tunneling electron using an effective one-electron potential such as that used in density functional theory would predict almost zero resonant current because of the exponentially small penetration of the bound state wave function under the triangular potential barrier of an applied electric field for sufficiently long molecules. Obviously, this conclusion contradicts experiment and requires new, non-standard approaches for investigation of the resonant transitions in single molecular systems.

We have recently developed a theory of electron transfer in single molecular systems which is based on Green's function theory of electron sub-barrier scattering off the structural molecular units (or functional groups) of a molecular chain [13,14]. The fundamental building block of this theory is the scattering operator $t_i(\varepsilon, \theta)$ of the structural unit or scattering center $i$. The scattering operator $t_i(\varepsilon, \theta)$ is calculated using a variational-asymptotic approach that explicitly takes into account the electron-electron and electron-nuclei interactions between the tunneling electron and the molecule [15]. All the complexity of the many-body interactions between the tunneling electron and the molecule is coarse grained into the energy $\varepsilon$ and the scattering angle $\theta$ dependence of the scattering operator. Once the $t_i(\varepsilon, \theta)$ are known for all structural units of the molecule, the total Green's function $G$ of the tunneling electron is easily obtained. The poles of $G$ comprise the bound energy spectrum of the tunneling electron, the energies and the corresponding wave-functions being obtained in the course of solving a system of linear equations for the partial scattering operators $\vec{T}(s) = \{T_1(s), T_2(s), ..., T_N(s)\}$ [13,14].

The effects of an external electric field are explicitly taken into account by parametric referencing of the local vacuum levels of the scattering centers by the local electrostatic energy $V_i = -V \dfrac{R_i}{d}$. The wave function for the bound state $s$ that determines the probabilities of $\gamma_l$ and $\gamma_r$ is obtained as

$$\psi_s(\mathbf{r}; \varepsilon_s) = \sum_n T_n(s) \tilde{\varphi}(\mathbf{r} - \mathbf{R}_n; \varepsilon_s), \qquad (3)$$

where $\tilde{\varphi}(\mathbf{r} - \mathbf{R}_n)$ are additional contributions to the exponential tail of the electron wave function of the tunneling electron due to its interaction with scattering center $n$. The partial scattering operators $T_n(s)$ are determined via solution of the homogeneous system of linear equations:

$$T_n(\varepsilon,V) = \sum_{k \neq n} t_n(\varepsilon - V_n) G_V(R_n, R_k; \varepsilon) T_k(\varepsilon,V), \qquad (4)$$

where $G_V(R_n, R_k; \varepsilon)$ is the vacuum Green's function in the external electric potential $V$ that connects centers $R_n$ and $R_k$,

$$G_V(\mathbf{R}_n, \mathbf{R}_k; \varepsilon) = -\frac{1}{2\pi |\mathbf{R}_n - \mathbf{R}_k|} \cdot \exp\left(-\frac{2\sqrt{2}|\mathbf{R}_l - \mathbf{R}_r|}{3V}\left\{(|\varepsilon|-|V_n|)^{3/2} - (|\varepsilon|-|V_k|)^{3/2}\right\}\right), \qquad (5)$$

The determinant of system (4) is the secular equation for the eigenspectrum of the bound states

$$\det|\delta_{ik} - t_i(\varepsilon - V_i, \vartheta) G_V(R_i, R_k; \varepsilon)| = 0. \qquad (6)$$

As is seen from (4)-(6), the quantity $h(\varepsilon, V_i) = |t_i(\varepsilon - V_i, \vartheta) G_V(R_i, R_{i\pm 1}; \varepsilon)|$ serves as an effective hopping integral that can be used to relate the properties of individual scattering centers to the local properties of the wave function of the resonant state of the tunneling electron $\psi_b(\varepsilon_s)$ as a function of applied bias. The effective hopping integrals $h(\varepsilon, V_i)$ determines the position of the lowest energy levels with respect to $E_F$ (and the value of the threshold voltage) as well as the values of the bound state wave function $\psi_b(\varepsilon_s, \mathbf{R}_{l,r})$ at the ends of the molecule.

Qualitatively, the larger the hopping integral $h(\varepsilon, V_i)$, the stronger the localization of the tunneling electron at center $i$. The effective hopping integral $h(\varepsilon, V_i)$ is a decreasing function of the absolute energy $|\varepsilon|$ due to an exponential reduction of the Green's function (5).

Therefore, when the positive bias is applied, the right end of the molecule experiences an effective reduction of the local vacuum level $\varepsilon - V$ ($\varepsilon, V < 0$) and the wave function is strongly localized at the right end of the molecule. At the left end of the molecule, electrostatic potential and a corresponding reduction of the local vacuum level are small which results in a relatively small value of the wave-function at the left end of the molecule $\psi_b(\mathbf{R}_l, \varepsilon_s)$. Therefore, it is the value of $\psi_b(\mathbf{R}_l, \varepsilon_s)$ at the left end that determines the contribution of the bound state $s$ to the resonant current via the probability $\gamma_l$, see (2). When the polarity of bias is reversed, localization occurs at the left end of the molecule which experiences a larger reduction of the local vacuum level. Then, the resonant current is determined by $\psi_b(\mathbf{R}_r, \varepsilon_s)$ at the right end of the molecule.

The energy dependence of the scattering operators $t_i(\varepsilon)$ is the key factor in explaining the physics of the rectification effect. In general, when the energy of a tunneling electron decreases with respect to the vacuum level ($|\varepsilon|$ increases), $|t_i(\varepsilon)|$ grows substantially due to an appreciable contribution to the exchange interaction of the tunneling electron with the electrons of the molecule at tunneling energies close to the energies of the molecular orbitals. Let us assume that the left half of the molecule consists of scattering centers which have larger values of the scattering operator $t_i(\varepsilon)$ than scattering centers on the right half of the molecule, $|t_l(\varepsilon)| > |t_r(\varepsilon)|$. This difference is due to different number of valence electrons in thiazole and thiophene structural units. At zero bias $|h_l(\varepsilon_s,0)/h_r(\varepsilon_s,0)| = |t_l(\varepsilon_s)/t_r(\varepsilon_s)| > 1$, consequently, $|\psi_b(\varepsilon_s,\mathbf{R}_l)| > |\psi_b(\varepsilon_s,\mathbf{R}_r)|$. As the bias is increased, this ratio is reduced due to the trend for wave function localization at the right end, by a rate which is determined by the ratio $|h_l(\varepsilon_s,V)/h_r(\varepsilon_s,V)|$. Eventually, this ratio becomes less than 1 at any bias above the threshold value, but the localization of the bound state of the tunneling electron at the right end of the molecule is greatly reduced because of the initial dominance of the scattering operators at the left end. Therefore, the probability $\gamma_l$ will be appreciable and the direct current is substantial.

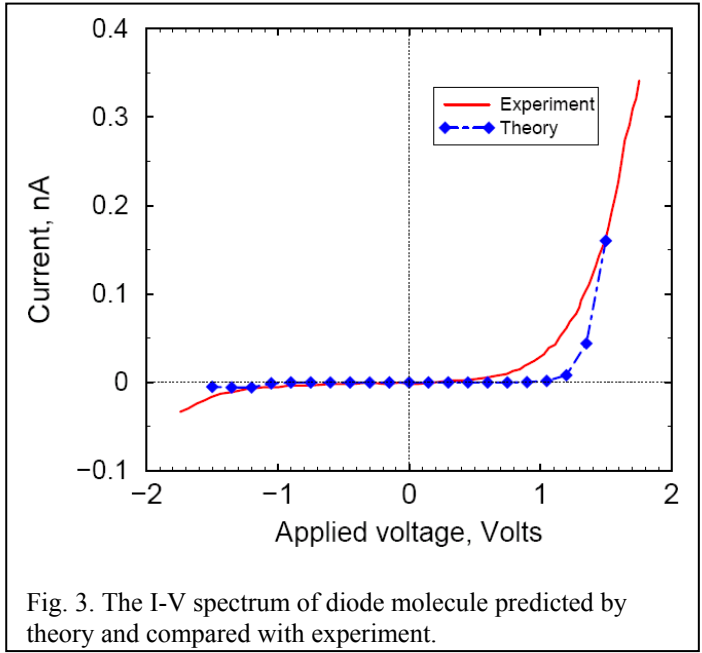

Fig. 3. The I-V spectrum of diode molecule predicted by theory and compared with experiment.

In the case of negative bias applied to the molecule, the localization at the left end of the molecule is greatly assisted due to the initial imbalance $|\psi_b(\varepsilon_s,\mathbf{R}_l)| > |\psi_b(\varepsilon_s,\mathbf{R}_r)|$. As bias is increased, localization at the left end occurs at a much higher rate due to the increased energy dependence of the scattering operator $t_l(\varepsilon_s)$. As a consequence, the wave function $\psi_b(\varepsilon_s,\mathbf{R}_r)$ at the right end at negative bias is much smaller than $\psi_b(\varepsilon_s,\mathbf{R}_l)$ at the left end at the same positive bias. Therefore, due to the much smaller value of $\gamma_r$, $\gamma_r < \gamma_l$, the resonant current at negative bias above the threshold will be much smaller that that at the same bias but of positive polarity, see eq. (2).

In order to confirm this qualitative picture, we have performed calculations of diode I-V characteristics for a di-block oligomer molecule consisting of four thiophene and four thiazole units. Each ring is modeled as a single structural unit, i.e. scattering center, and is represented by the scattering operators $t_1(\varepsilon)$ or $t_2(\varepsilon)$ for thiazole and thiophene rings. The parametrization of $t_1(\varepsilon)$ or $t_2(\varepsilon)$ used in simulations is

$$t(\varepsilon)/2\pi = a/(\varepsilon - \varepsilon_0) + b + c\varepsilon, \qquad (7)$$

where $a/(\varepsilon - \varepsilon_0)$ and $b + c\varepsilon$ are the pole and potential terms of the scattering operator respectively. The numerical values of $a_1 = a_2 = 1.0$ $\varepsilon_{0,1} = \varepsilon_{0,2} = -0.037$; $b_1 = b_2 = 10.0$; $c_1 = 2500$; $c_2 = 1500$ were chosen based on our previous calculations of scattering operators for various molecular systems and taking into account the specific conditions of the experiment. In particular, the values of the coefficients in (7) are an order of magnitude larger than that for simple systems such as the hydrogen atom, which is quite reasonable when taking into account the fact that the scatterers are molecular systems consisting of several atoms connected by conjugated bonds. The distance $d$ between the scatterers is $4 \overset{\circ}{\text{A}} \approx 8$ *a.u.*

The calculated I-V spectrum for a molecular diode is shown and compared with experiment in Fig. 3. This simple model of a molecular diode predicts threshold values of the voltage $V_l^{th} \approx 1.0 \text{ eV}$ and a rectification ratio $I(+1.5\text{V})/I(-1.5\text{V}) = 10$ that is in qualitative agreement with experiment.

As stressed above, the rectification effect in diode molecules is the consequence of resonant electron transfer in di-block oligomer systems. Interestingly, a similar asymmetry in the I-V spectrum with respect to the polarity of applied bias can be observed in the regime of direct tunneling through single molecules, i.e. when the energy levels of the tunneling electron are above the Fermi level of the negatively biased electrode. However, the rectification ratio and absolute values of the tunneling current are much smaller than in the resonant case.

In conclusion, we investigated the mechanism of rectification in di-block oligomer diode molecules and have shown that the observed rectification effect is due to both the resonant nature of electron transfer in the system and the localization properties of bound state wave functions of the resonant states of the tunneling electron in an electric field. We have shown, that the standard concepts of quantum chemistry and solid state physics such as HOMO-LUMO and p-type and n-type conduction that are usually used to interpret experimental results are not applicable in the case of single electron transport in molecular systems. This conceptually new theoretical approach, the Green's function theory of sub-barrier scattering, provides a physically transparent explanation of the rectification effect based on the concept of the bound state spectrum of a tunneling electron and predicts the characteristic features of an I-V spectrum in qualitative agreement with experiment.

M.A.K. and V.S.P. thank the Russian Foundation for Basic Research for financial support under grant 02-03-33028. I.I.O. thanks the National Science Foundation for financial support under grant CCF-0432121. L.P. Yu thanks the National Science Foundation for financial support under grants DMR-0213745 (MRSEC) and DMR-0139155

**REFERENCES**
1. A. Aviram, and M.A. Ratner, Chem. Phys. Lett. **29**, 277 (1974).
2. A.S. Martin, J.R. Sables, and G.J. Ashwell, Phys. Rev. Lett. **70**, 218 (1993).
3. M. Pomerantz, A. Aviram, R.A. McCorkle, L. Li, and A.G. Schrott, Science **255**, 1115 (1992).
4. C.M. Fischer, M. Burghard, S. Roth, K.V. Klitzing, Europhys. Lett. **28**, 129 (1994)
5. C.P. Collier, E. W. Wong, M. Belohradsky, F. M. Raymo, J. F. Stoddart, P. J. Kuekes, R. S. Williams, and J. R. Heath, Science **285**, 391 (1999).
6. R.M. Metzger, B. Chen, U. Höpfner, *et al,* J. Am. Chem. Soc. **119,** 10455 (1997).
7. For review, see R.M. Metzger, Ann. N.Y. Acad. Sci. **1006**, 252 (2003).
8. M.K. Ng, and L.P. Yu, Angew. Chem. Engl. Ed **41,** 3598 (2002).
9. M.K. Ng, D.C. Lee, and L.P. Yu, J. Am. Chem. Soc., **124,** 11862 (2002).
10. P. Jiang, G.M. Morales, W. You, L.P Yu, Angew. Chem. Engl. Ed. **43**, 4471 (2004).
11. M.A. Kozhushner, V.S. Posvyanskii and I. I. Oleynik, submitted for publication (2005).
12. M.A. Kozhushner, V.S. Posvyanskii and I. I. Oleynik, submitted for publication (2005).
13. D.I. Bolgov, M.A. Kozhushner, R.R. Muryasov, and V.S. Posvyanskii, J. Chem. Phys. **119**, 3871 (2003).
14. M.A. Kozhushner, V.S. Posvyanskii and I. I. Oleynik, submitted for publication (2004).
15. M.A. Kozhushner and R.R. Muriasov, Dokl. Phys. Chem. **372**, 82 (2000).